# *An absolute sodium abundance for a cloud-free 'hot Saturn' exoplanet*

*Nikolay Nikolov[1*], David K. Sing[1,2], Jonathan J. Fortney[3], Jayesh M. Goyal[1], Benjamin Drummond[1], Tom M. Evans[1], Neale P. Gibson[4], Ernst J. W. De Mooij[5,6], Zafar Rustamkulov[3], Hannah R. Wakeford[7], Barry Smalley[8], Adam J. Burgasser[9], Coel Hellier[8], Christiane Helling[10,11], Nathan J. Mayne[1], Nikku Madhusudhan[12], Tiffany Kataria[13], Josef Baines[4], Aarynn L. Carter[1], Gilda E. Ballester[14], Joanna K. Barstow[15], Jack McCleery[4] & Jessica J. Spake[1]*

*[1] Physics and Astronomy, University of Exeter, Exeter EX4 4QL, UK.*
*[2] Department of Earth and Planetary Sciences, Johns Hopkins University, Baltimore, MD, USA*
*[3] Department of Astronomy and Astrophysics, University of California, Santa Cruz, CA 95064, USA.*
*[4] Astrophysics Research Centre, School of Mathematics and Physics, Queens University Belfast, Belfast BT7 1NN, UK.*
*[5] School of Physical Sciences, and Centre for Astrophysics & Relativity, Dublin City University, Glasnevin, Dublin 9, Ireland.*
*[6] Centre for Astrophysics & Relativity, Dublin City University, Glasnevin, Ireland*
*[7] Space Telescope Science Institute, 3700 San Martin Drive, Baltimore, Maryland 21218, USA.*
*[8] Astrophysics Group, Keele University, Keele, UK.*
*[9] Department of Physics, University of California, San Diego, CA 92093, USA..*
*[10] Centre for Exoplanet Science, SUPA, School of Physics and Astronomy, University of St. Andrews, North Haugh, St. Andrews, Fife, KY16 9SS, UK.*
*[11] Anton Pannekoek Institute for Astronomy, University of Amsterdam, Science Park 904, 1098 XH Amsterdam, The Netherlands.*
*[12] Institute of Astronomy, University of Cambridge, Madingley Road, CB3 0HA Cambridge, UK.*
*[13] Jet Propulsion Laboratory, California Institute of Technology, 4800 Oak Grove Drive, Pasadena, CA, USA.*
*[14] Lunar and Planetary Laboratory, University of Arizona, Tucson, AZ 85721, USA.*
*[15] Physics and Astronomy, University College London, London, UK.*
** e-mail: nikolov.nkn@gmail.com*

**Broad absorption signatures from alkali metals, such as the sodium (Na I) and potassium (K I) resonance doublets, have long been predicted in the optical atmospheric spectra of cloud-free irradiated gas-giant exoplanets[1,2,3]. However, observations have only revealed the narrow cores of these features rather than the full pressure-broadened profiles[4-6]. Cloud and haze opacity at the day-night planetary terminator are considered responsible for obscuring the absorption-line wings, which hinders constraints on absolute atmospheric abundances[7-9]. Here we present an optical transmission spectrum for the 'hot-Saturn' WASP-96b obtained with the Very Large Telescope, which exhibits the complete pressure-broadened profile of the sodium absorption feature. The spectrum is in excellent agreement with cloud-free, solar-abundance models assuming chemical equilibrium. We are able to measure a precise, absolute sodium abundance of log $\varepsilon_{Na} = 6.9^{+0.6}_{-0.4}$, and use it as a proxy to the planet's atmospheric metallicity relative to the solar value ($Z_p/Z_\odot = 2.3^{+8.9}_{-1.7}$). This result is consistent with the mass-metallicity trend observed for solar-system planets and exoplanets[10-12].**

We observed two transits of the hot Saturn WASP-96b (planetary mass $M_\mathrm{p} = (0.48 \pm 0.03)\ M_\mathrm{J}$, $R_\mathrm{p} = (1.20 \pm 0.06)\ R_\mathrm{J}$, $T_\mathrm{eq} = 1285 \pm 40\mathrm{K}$)[13] on 2017 July 29th and August 22nd UT in photometric conditions, using the 8.2-metre Unit Telescope 1 of the Very Large Telescope, with the FORS2 spectrograph. Data were collected in the multi-object-spectroscopy mode using grisms 600B (blue) and 600RI (red) on the first and second nights, respectively, which combined cover the wavelength range 3600 to 8200 Å. We used a mask consisting of two broad slits centred on the target and on a reference star of similar brightness. Broad slits spanning 22″ along the dispersion and 120″ along the spatial (perpendicular) axis were used in order to minimise slit losses due to seeing variations and guiding imperfections.

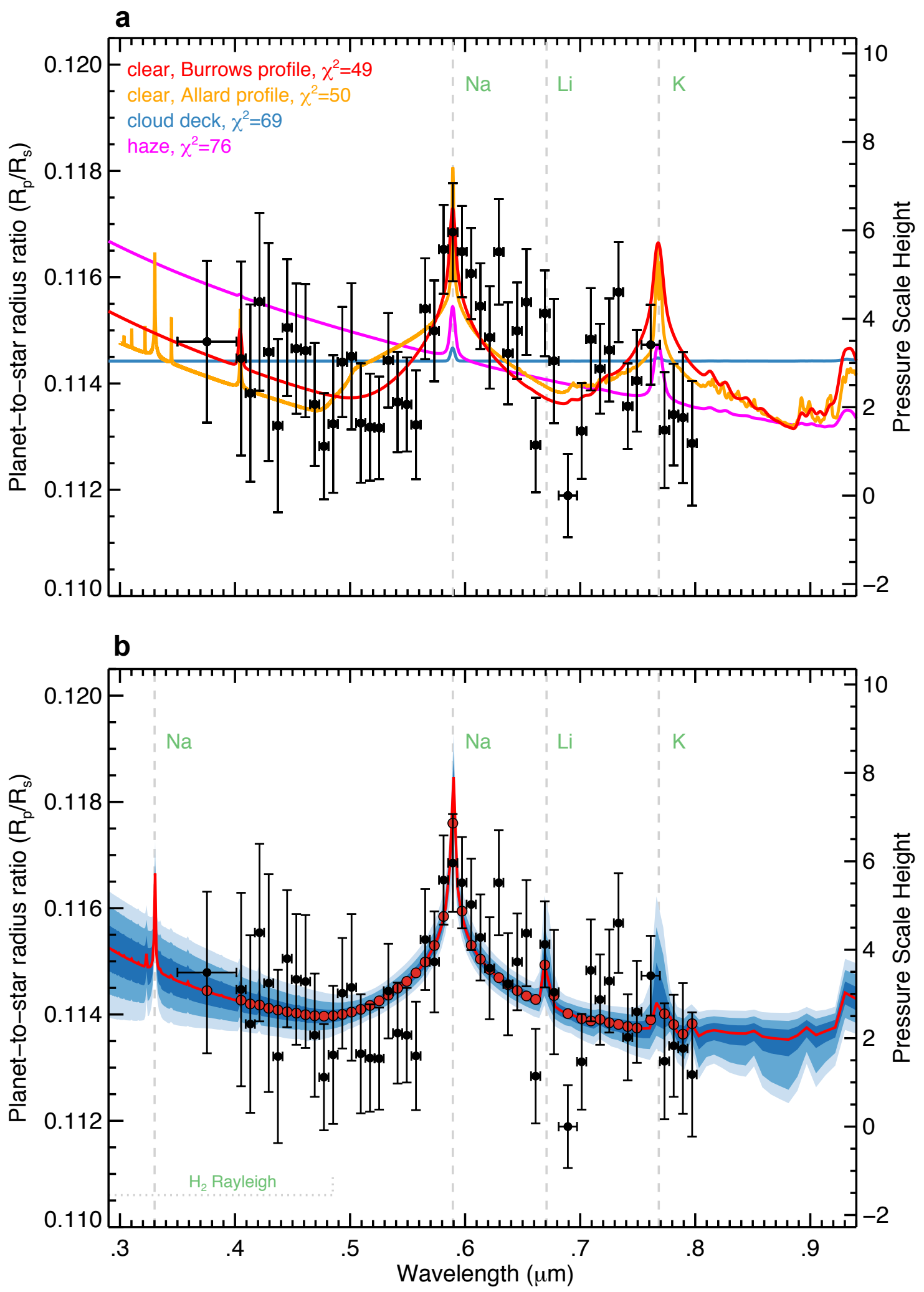

**Figure 1 | Transmission spectrum of WASP-96b compared to models. a:** Comparison of the FORS2 observations (black dots with 1$\sigma$ vertical error bars; the horizontal bars indicate spectral bin widths) with clear[3,16], cloudy and hazy one-dimensional forward atmospheric models at solar abundance[14] (continuous lines). The two best-fit models assume a clear atmosphere with different line broadening shapes for Na and K (see text for details). Models with hazes or clouds (magenta and blue) predict much smaller and narrower absorption features. **b:** Similar to the upper panel, but showing the best-fit model obtained from the retrieval analysis (red line) binned to the data resolution (red dots), with the 1$\sigma$, 2$\sigma$ and 3$\sigma$ confidence intervals (blue regions).

For each transit, we produced wavelength-integrated "white" and spectroscopic light curves for WASP-96 and the reference star by integrating the flux of each spectrum along the dispersion axis. We corrected the light curves for extinction caused by the Earth's atmosphere by dividing the flux of the target by the flux of the reference star. We modelled the transit and systematic effects of the white-light curves by treating the data as a Gaussian Process (GP) and assuming quadratic limb darkening for the star. The transit parameters: mid-time ($T_\mathrm{mid}$), orbital inclination ($i$), normalised semi-major axis ($a/R_*$), planet-to-star radius ratio ($R_p/R_*$) and the two limb-darkening coefficients ($u_1$ and $u_2$), were allowed to vary in the fit to each of the two white-light curves, while the orbital period was held fixed to the previously determined value. The white-light curves and results from the modelling are shown in Extended Data Figure 1 and Table 1.

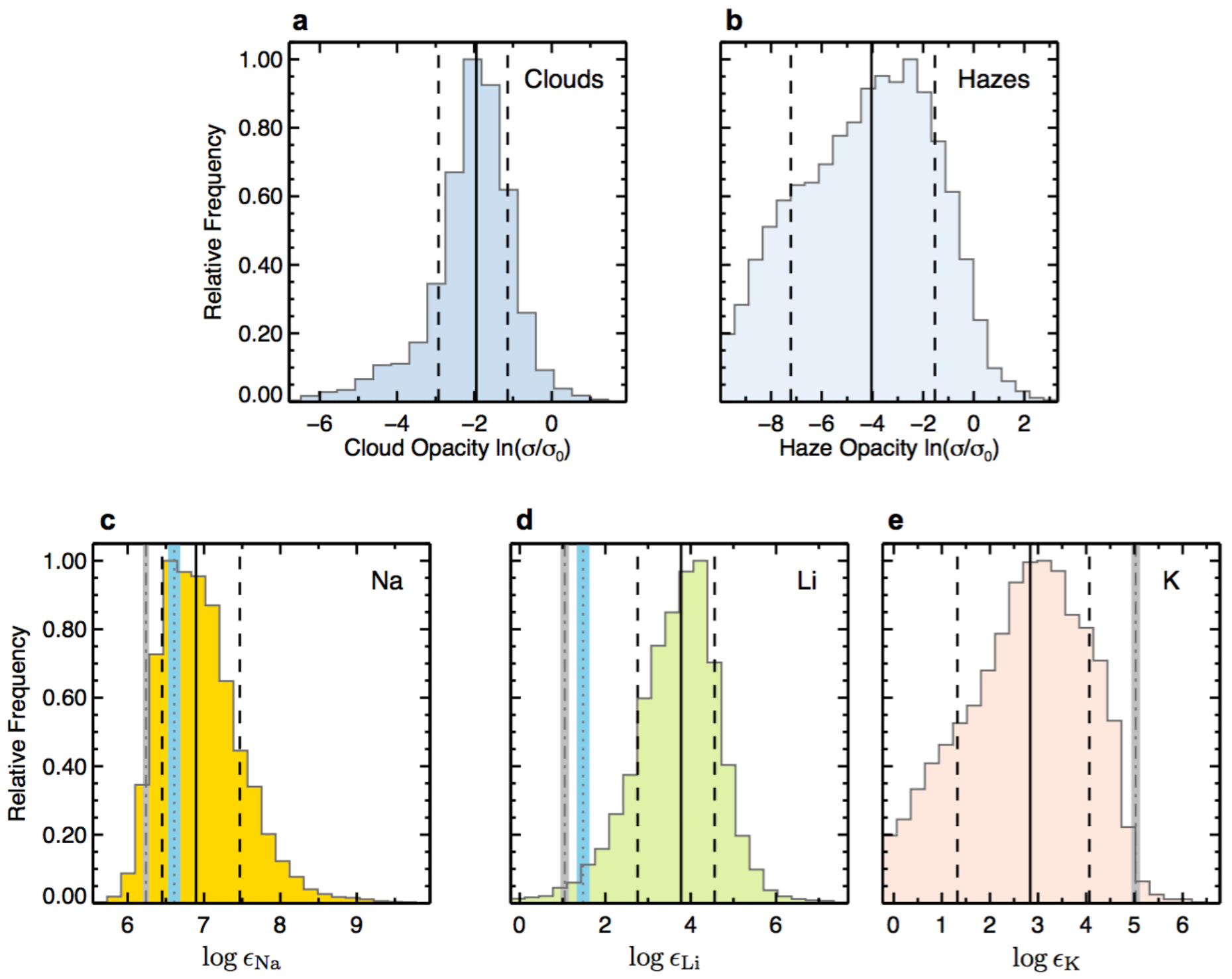

**Figure 2 | Retrieved atmospheric properties for WASP-96b.** Histograms of the marginalized posterior distributions from a free retrieval. **a, b,** Negligible opacity from clouds and hazes comprises the evidence for a clear atmosphere at the limb of the planet. **c-e,** Retrieved elemental abundances in the scale of ref.[23], which ranges from 0 to 12 with the abundance of hydrogen $\log \varepsilon_{\mathrm{H}} = 12$. The abundance of Na, $\log \varepsilon_{\mathrm{Na}} = 6.9^{+0.6}_{-0.4}$ is the only constrained quantity (**c**). The vertical continuous and dotted lines indicate the mean abundances and $1\sigma$ uncertainties, respectively. Shown are the elemental abundances with the uncertainties of the host star (dotted lines and blue regions) and the Sun (dash-dotted lines in grey regions from ref.[25]).

To obtain the transmission spectrum, we produced 28 and 35 spectroscopic light curves, from the blue and red grisms, respectively, with a width of 160 Å. Wavelength-independent systematics were corrected following standard practices, as detailed in Methods. We allowed only $R_p/R_*$ and $u_1$ to vary. The rest of the system parameters were fixed to their weighted mean values from the analysis of the two white-light curves, while the quadratic limb-darkening coefficients ($u_2$) were fixed to their theoretical values. To account for systematics, we marginalised over a grid of polynomials, where the latter consisted of terms up to second order in air mass and drift of the spectrum across the detector along both the dispersion and cross-dispersion axes. The resulting time series are shown in Extended Data Figure 2 and 3.

The measured wavelength-dependent relative planet radii are shown in Figure 1, which comprises the transmission spectrum of WASP-96b. The spectrum reveals the absorption signature of the pressure-broadened sodium D line with wings covering ~6 atmospheric pressure scale heights (1 scale height corresponds to ~610 km, assuming $T_{eq}$ = 1285K), in a wavelength range from ~5000 to ~7500 Å, and a slope at near-UV wavelengths due to Rayleigh scattering by molecular hydrogen. The radius measurements around the potassium feature show no obvious broadened line wing shape or larger absorption at the line cores.

To interpret the measured transmission spectrum, we first compare it with clear, cloudy and hazy atmospheric models with solar abundances from ref. 14. We find that cloud-free models assuming chemical equilibrium best fitted the 49 data points, giving $\chi^2 = 49$ and $\chi^2 = 50$ for a total of 48 degrees of freedom. Models with clouds and hazes, i.e. 100× enhanced-Rayleigh scattering cross-section (haze) and 100× enhanced wavelength-independent (cloud) opacity, give $\chi^2$ of 69 and 76 respectively, and are disfavoured at ~3$\sigma$ and ~5$\sigma$ confidence, respectively (Figure 1). Further details are provided in Methods.

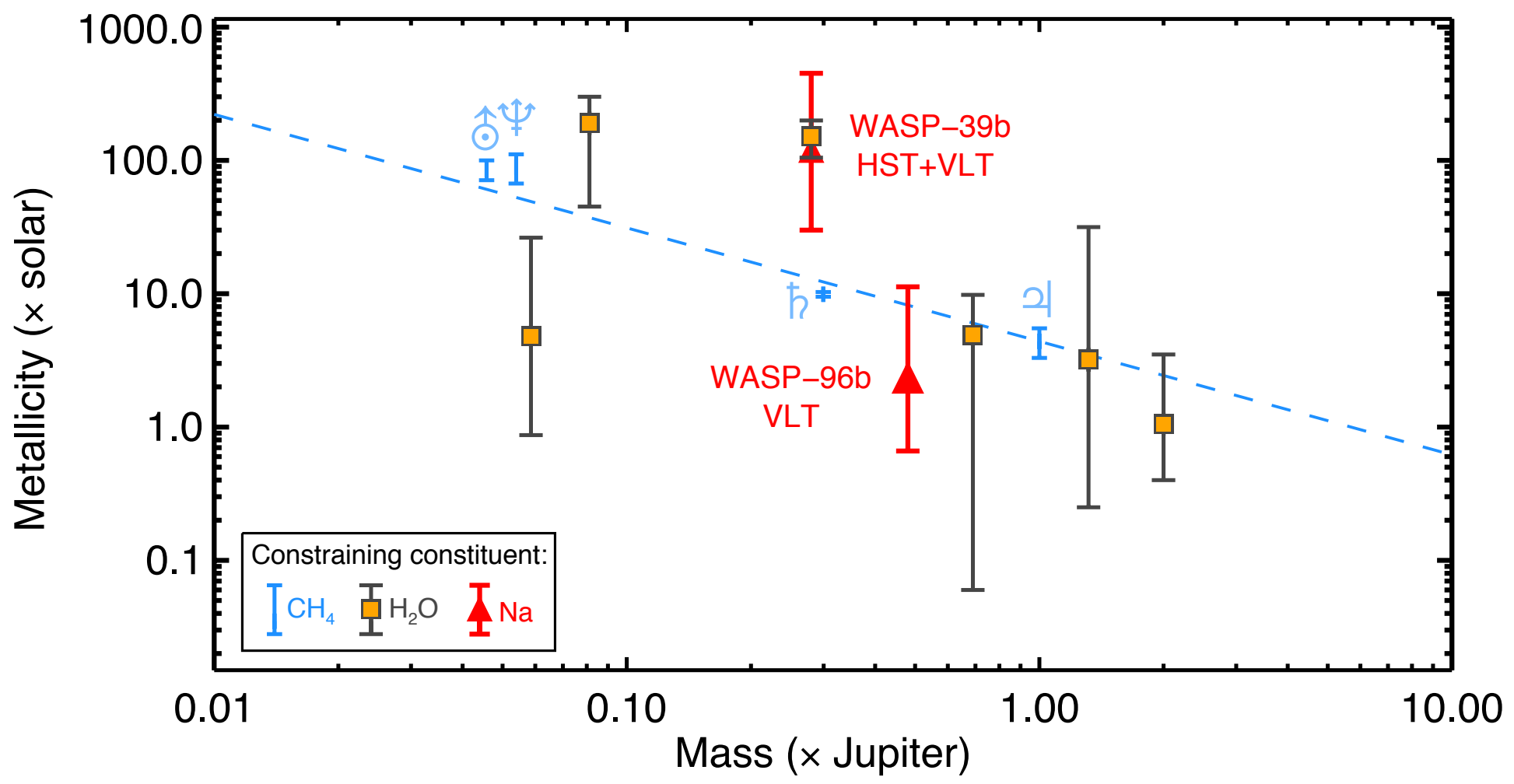

**Figure 3 | Mass-metallicity diagram for Solar System planets and exoplanets.** Methane ($CH_4$) and water ($H_2O$) are the two absorbing constituents used to constrain the atmospheric metallicity of solar system planets (blue bars) and hot gas-giant exoplanets (orange squares with grey error bars), respectively. Absorption lines from atomic Na (red triangles) can provide another proxy to exoplanet atmospheric metallicity, which has been done by combining three Hubble Space Telescope (HST) and two Very Large Telescope (VLT) transits for WASP-39b. With its detected and resolved pressure-broadened Na line wings, WASP-96b is the first transiting exoplanet for which high-precision atmospheric metallicity has been constrained using data only from the ground. Each error bar corresponds to the 1$\sigma$ uncertainty. The blue line indicates a fit to the Solar System gas giants (pale blue symbols indicate Solar System planets).

The wing shape of atomic absorption lines is a result of the combined contribution of the quantum mechanical (natural), thermal (or Doppler) and collisional (or pressure) broadening mechanisms[15]. Measurements of the shape of pressure-broadened line wings can provide important constraints on the interaction potentials used in the theory of stellar and sub-stellar atmospheres[16,17]. While, such constraints have been obtained from Na and K absorption lines in the spectra of brown dwarfs[18,19], the actual shape of the profiles for exoplanets remain unconstrained. To assess the detection of sodium line-broadening we compared the spectrum to models with no broadened lines. Compared to the best-fit clear-atmosphere model, the narrow-line model is found to be rejected at the $5.8\sigma$ confidence level. This is in contrast to WASP-39b, WASP-17b and HD209458b, which have previously been classified as having the clearest atmospheres of the known exoplanets. The latter transmission spectra are well-explained with narrow alkali features, implying that the broad absorption wings are masked by clouds and hazes[4,5,20,21].

The broad sodium feature measured for WASP-96b therefore provides a unique opportunity to constrain the pressure-broadened line shape for an exoplanet atmosphere. We compared the observed spectrum to two cloud-free models, assuming alkali line-wing shapes from refs. 3 and 16. We find each of them to be statistically consistent with the data, although the wing profile of ref. 3 is marginally preferred (Figure 1, red and orange models).

To further interpret the physical properties of WASP-96b's atmosphere, we performed a retrieval analysis of the data using the 1D radiative-convective ATMO model[22]. We assumed an isothermal atmosphere and allowed the temperature, radius, opacity from clouds and hazes and elemental abundances of Na and K to vary. In addition, Li is expected to add opacity at $\sim 6650$ Å, which is covered by three of our measurements. Throughout this letter, we adopt the astronomical scale of logarithmic abundances of ref. 23, where hydrogen (H) is defined to be $log\ \varepsilon_{\mathrm{H}} = 12$. The abundance of a particular element X is defined as $\log \varepsilon_{\mathrm{X}} = \log(N_{\mathrm{X}}/N_{\mathrm{H}}) + 12$, where $N_{\mathrm{X}}$ and $N_{\mathrm{H}}$ are the number densities of elements X and H. Our retrieval analysis finds negligible contributions from cloud or haze opacity, which indicates that the atmosphere of WASP-96b is free of clouds and hazes at the pressures being probed at the limb. The best-fit transmission spectrum includes opacity from Na, Li, K and Rayleigh scattering (Figure 1). We obtain a tight constrain of $\log\ \epsilon_{\mathrm{Na}} = 6.9^{+0.6}_{-0.4}$ on the sodium abundance, which is in agreement with the solar abundance as well as with the measured sodium abundance in the WASP-96 host star (Figure 2). The best-fit model gives $\chi^2 = 39$ for 42 degrees of freedom.

The current data does not support detections of K or Li, as the minimum $\chi^2$ value when excluding the two species is only slightly higher than when they are included ($\Delta\chi^2 = 2$). However, we include the two species in our retrieval model to marginalise the Na abundance over the possibility of their presence and estimate upper-limits on their abundances. The abundances of K and Li are also found to depend on the assumed profile shape of the Na feature. We find an atmospheric temperature of $\mathrm{T} = 1710^{+150}_{-200}$K, which is somewhat higher, compared with the planet's equilibrium temperature of $\mathrm{T_{eq}} = 1285 \pm 40$K under the assumption of zero albedo and uniform day-night heat redistribution [13].

**Extended Data Table 1 | System parameters**

| Parameter | Value |
|---|---|
| Orbital period (day) | 3.4252602 (fixed) |
| eccentricity | 0 (fixed) |
| | |
| GRIS600B | 4013 – 6173 Å |
| $T_{mid}$ (JD) | $57963.33672^{+0.00034}_{-0.00033}$ |
| i (°) | $85.21^{+0.18}_{-0.17}$ |
| $a/R_*$ | $8.93^{+0.18}_{-0.18}$ |
| $R_p/R_*$ | $0.1141^{+0.0015}_{-0.0014}$ |
| $u_1$ | $0.399^{+0.056}_{-0.030}$ |
| $u_2$ | $0.148^{+0.075}_{-0.040}$ |
| A (ppm) | $501^{+107}_{-72}$ |
| $\eta_{FWHM}$ (arbitrary normalisation) | $1.01^{+1.2}_{-0.99}$ |
| $\eta_X$ (arbitrary normalisation) | $-3.45^{+0.70}_{-0.57}$ |
| $\eta_Y$ (arbitrary normalisation) | $-1.86^{+0.88}_{-1.01}$ |
| $c_0$ | $0.9992^{+0.00017}_{-0.00016}$ |
| $c_1$ | $-0.00068^{+0.00012}_{-0.00012}$ |
| | |
| GRIS600RI | 5268 – 8308 Å |
| $T_{mid}$ (JD) | $57987.31195^{+0.00029}_{-0.00029}$ |
| i (°) | $85.11^{+0.12}_{-0.12}$ |
| $a/R_*$ | $8.80^{+0.11}_{-0.12}$ |
| $R_p/R_*$ | $0.1172^{+0.0017}_{-0.0017}$ |
| $u_1$ | $0.26^{+0.11}_{-0.08}$ |
| $u_2$ | $0.23^{+0.12}_{-0.09}$ |
| A (ppm) | $2301^{+1690}_{-741}$ |
| $\eta_{RAS}$ (arbitrary normalisation) | $0.50^{+0.87}_{-0.69}$ |
| $\eta_X$ (arbitrary normalisation) | $2.81^{+0.94}_{-0.76}$ |
| $\eta_Y$ (arbitrary normalisation) | $8.08^{+3.5}_{-2.5}$ |
| $c_0$ | $0.9980^{+0.0016}_{-0.0013}$ |
| $c_1$ | $0.00018^{+0.00029}_{-0.00028}$ |
| | |
| Weighted mean: | |
| i (°) | $85.14 \pm 0.10$ |
| $a/R_*$ | $8.84^{+0.09}_{-0.10}$ |
| GRIS600B | (fixed i, $a/R_*$ and $T_{mid}$) |
| $R_p/R_*$ | $0.1147^{+0.0014}_{-0.0014}$ |
| $u_1$ | $0.435^{+0.035}_{-0.020}$ |
| $u_2$ | $0.173^{+0.041}_{-0.022}$ |
| GRIS600RI | (fixed i, $a/R_*$ and $T_{mid}$) |
| $R_p/R_*$ | $0.1168^{+0.0015}_{-0.0014}$ |
| $u_1$ | $0.282^{+0.053}_{-0.032}$ |
| $u_2$ | $0.250^{+0.053}_{-0.032}$ |

Heavy element abundance measurements are important to constrain formation mechanisms of gas-giant exoplanets. Under the core-accretion paradigm, as the planet mass decreases the atmospheric metallicity increases[24,25]. As giant planets accrete H/He-dominated gas as they form, they also accrete planetesimals[26] that enrich their H/He envelopes in metals. A low-mass H/He envelope has smaller amount of gas for these metals to be mixed into, leading to a higher metal enrichment compared to the parent star. This is also the scenario for solar system gas-giants, where metallicity has been constrained from methane ($CH_4$) abundance from in-situ or infrared-spectroscopy[27-30], showing increasing enrichment of heavy elements with decreasing mass (Figure 3). Measurements of $H_2O$ abundances have been used to constrain atmospheric metallicities for a small sample of exoplanets[10-12]. The measured molecular abundances are used as proxies to atmospheric metallicities, assuming chemical equilibrium conditions. Using our measurement of the absolute sodium abundance of WASP-96b, we estimate an atmospheric metallicity of $Z_{\mathrm{p}}/Z_{\odot} = 2.3^{+8.9}_{-1.7}$, i.e. $\log{(Z_{\mathrm{p}}/Z_{\odot})} = 0.4^{+0.7}_{-0.5}$. This is consistent with the heavy element abundance of the host star $Z_{*}/Z_{\odot} = 1.4 \pm 0.7$ which we estimate using the relation $Z_{*}/Z_{\odot} = 10^{[\mathrm{Fe/H}]}$, where $[\mathrm{Fe/H}] = 0.14 \pm 0.19$. While our WASP-96b measurement is consistent with the solar system mass-metallicity trend (see Figure 3), we note that additional high-precision constraints would be necessary to further support or refute a trend for exoplanets.

WASP-96b is the first exoplanet for which the pressure-broadened wings of an atomic absorption line (Na I) have been observed, probing deeper layers of the atmosphere at the limb. This observation has also enabled a precise atmospheric abundance constraint, using ground-based data alone. Our result demonstrates that combined with near-UV data, the $\sim$5890Å Na absorption feature is a valuable probe of exoplanet metallicities accessible to ground-based telescopes over a wavelength region largely free of contamination by telluric lines. WASP-96b is the first gas-giant with a detected broad atomic absorption feature out of approximately 20 exoplanets so far characterized in transmission. This demonstrates the significant role a future ground-based optical spectrograph, optimized for transmission spectroscopy, could play. With the clearest atmosphere of any exoplanet characterized so far, WASP-96b will be an important target for the upcoming James Webb Space Telescope.

**Online Content** Any Methods, including any statements of data availability and Nature Research reporting summaries, along with any additional references and Source Data files, are available in the online version of the paper at https://doi.org/10.1038/s41586-018-0101-7.

**Acknowledgements.** This work is based on observations collected at the European Organization for Astronomical Research in the Southern Hemisphere under European Southern Observatory programme 199.C-0467(H). The research leading to these results received funding from the European Research Council under the European Union's Seventh Framework Programme (FP7/2007-2013)/ERC grant agreement number 336792. A.J.B. is a US/UK Fulbright Scholar. J.M.G. and N.J.M. acknowledge support from a Leverhulme Trust Research Project Grant. J.K.B. is a Royal Astronomical Society Research Fellow.

**Author Contributions** N.N. led the design of the VLT FORS2 Large Programme and the scientific proposal (with contributions from N.P.G., D.K.S. and T.M.E.). N.N. led the observations, analysis, comparison with forward models and the interpretation of the result. D.K.S. led the atmospheric retrievals (with contributions from J.M.G.). J.J.F. and Z.R. provided forward atmospheric models for comparative analysis. B.S. and C.H. provided elemental abundances of the host star. N.N. wrote the manuscript (with contributions from D.K.S. and T.M.E.). J.B. and J.McC. performed independent tests on various parts of the data reduction and analysis as part of their final-year undergraduate projects under supervision from N.P.G. All authors discussed the results and commented on the draft.

**Author Information** Reprints and permissions information is available at www.nature.com/reprints. The authors declare no competing financial interests. Readers are welcome to comment on the online version of the paper. Correspondence and requests for materials should be addressed to nikolov.nkn@gmail.com

**Extended Data Table 2 | Transmission spectrum**

| $\lambda$ (Å) | $R_p/R_*$ | $u_1$ | $u_2$ |
|---|---|---|---|
| 3500 − 4013 | 0.11479 ± 0.00152 | 0.484 ± 0.059 | 0.207 |
| 4013 − 4093 | 0.11447 ± 0.00182 | 0.553 ± 0.059 | 0.219 |
| 4093 − 4173 | 0.11382 ± 0.00167 | 0.539 ± 0.053 | 0.223 |
| 4173 − 4253 | 0.11554 ± 0.00167 | 0.493 ± 0.052 | 0.227 |
| 4253 − 4333 | 0.11459 ± 0.00205 | 0.458 ± 0.054 | 0.231 |
| 4333 − 4413 | 0.11321 ± 0.00163 | 0.464 ± 0.049 | 0.235 |
| 4413 − 4493 | 0.11505 ± 0.00129 | 0.479 ± 0.043 | 0.239 |
| 4493 − 4573 | 0.11466 ± 0.00123 | 0.522 ± 0.039 | 0.242 |
| 4573 − 4653 | 0.11462 ± 0.00125 | 0.415 ± 0.043 | 0.246 |
| 4653 − 4733 | 0.11361 ± 0.00116 | 0.422 ± 0.042 | 0.251 |
| 4733 − 4813 | 0.11282 ± 0.00100 | 0.404 ± 0.038 | 0.255 |
| 4813 − 4893 | 0.11324 ± 0.00130 | 0.385 ± 0.040 | 0.256 |
| 4893 − 4973 | 0.11440 ± 0.00104 | 0.337 ± 0.041 | 0.258 |
| 4973 − 5053 | 0.11451 ± 0.00138 | 0.367 ± 0.044 | 0.262 |
| 5053 − 5133 | 0.11326 ± 0.00112 | 0.410 ± 0.038 | 0.269 |
| 5133 − 5213 | 0.11318 ± 0.00101 | 0.288 ± 0.043 | 0.273 |
| 5213 − 5293 | 0.11317 ± 0.00096 | 0.336 ± 0.040 | 0.275 |
| 5293 − 5373 | 0.11444 ± 0.00089 | 0.260 ± 0.039 | 0.277 |
| 5373 − 5453 | 0.11365 ± 0.00095 | 0.349 ± 0.047 | 0.278 |
| 5453 − 5533 | 0.11361 ± 0.00089 | 0.299 ± 0.042 | 0.283 |
| 5533 − 5613 | 0.11322 ± 0.00102 | 0.330 ± 0.046 | 0.287 |
| 5613 − 5693 | 0.11541 ± 0.00095 | 0.303 ± 0.050 | 0.290 |
| 5693 − 5773 | 0.11499 ± 0.00095 | 0.284 ± 0.044 | 0.291 |
| 5773 − 5853 | 0.11653 ± 0.00084 | 0.263 ± 0.044 | 0.295 |
| 5853 − 5933 | 0.11685 ± 0.00092 | 0.270 ± 0.041 | 0.296 |
| 5933 − 6013 | 0.11648 ± 0.00086 | 0.236 ± 0.049 | 0.296 |
| 6013 − 6093 | 0.11607 ± 0.00086 | 0.242 ± 0.050 | 0.297 |
| 6093 − 6173 | 0.11545 ± 0.00081 | 0.169 ± 0.053 | 0.302 |
| 6173 − 6253 | 0.11487 ± 0.00096 | 0.316 ± 0.044 | 0.313 |
| 6253 − 6333 | 0.11648 ± 0.00099 | 0.235 ± 0.041 | 0.314 |
| 6333 − 6413 | 0.11457 ± 0.00096 | 0.194 ± 0.045 | 0.314 |
| 6413 − 6493 | 0.11499 ± 0.00091 | 0.185 ± 0.046 | 0.314 |
| 6493 − 6573 | 0.11553 ± 0.00100 | 0.267 ± 0.043 | 0.313 |
| 6573 − 6653 | 0.11284 ± 0.00089 | 0.250 ± 0.040 | 0.313 |
| 6653 − 6733 | 0.11532 ± 0.00081 | 0.192 ± 0.038 | 0.313 |
| 6733 − 6813 | 0.11442 ± 0.00117 | 0.155 ± 0.044 | 0.314 |
| 6813 − 6973 | 0.11189 ± 0.00078 | 0.087 ± 0.042 | 0.314 |
| 6973 − 7053 | 0.11311 ± 0.00090 | 0.100 ± 0.052 | 0.314 |
| 7053 − 7133 | 0.11483 ± 0.00096 | 0.129 ± 0.051 | 0.314 |
| 7133 − 7213 | 0.11428 ± 0.00085 | 0.104 ± 0.044 | 0.315 |
| 7213 − 7293 | 0.11463 ± 0.00097 | 0.156 ± 0.048 | 0.315 |
| 7293 − 7373 | 0.11572 ± 0.00094 | 0.277 ± 0.038 | 0.315 |
| 7373 − 7453 | 0.11357 ± 0.00081 | 0.102 ± 0.050 | 0.316 |
| 7453 − 7533 | 0.11405 ± 0.00096 | 0.263 ± 0.046 | 0.316 |
| 7533 − 7693 | 0.11473 ± 0.00075 | 0.147 ± 0.042 | 0.315 |
| 7693 − 7773 | 0.11312 ± 0.00109 | 0.190 ± 0.049 | 0.315 |
| 7773 − 7853 | 0.11341 ± 0.00096 | 0.189 ± 0.049 | 0.315 |
| 7853 − 7933 | 0.11336 ± 0.00123 | 0.175 ± 0.053 | 0.314 |
| 7933 − 8013 | 0.11287 ± 0.00117 | 0.214 ± 0.059 | 0.314 |

# METHODS

## Observations.

We observed two transits of WASP-96b with the FOcal Reducer and Spectrograph (FORS2)[31] attached on the Unit Telescope 1 (UT1, Antu) of the Very Large Telescope (VLT) at the European Southern Observatory (ESO) on Cerro Paranal in Chile as part of Large Program 199.C-0467 (PI: Nikolov). We used similar observing setup and strategy to our VLT FORS2 Comparative Transmission Spectroscopy of WASP-39b and WASP-31b[22,32].

During the two transits, we monitored the flux of WASP-96 and one reference star at photometric conditions. The reference star, known as 2MASS 00041885-4716309 is the only bright source in the FORS2 field of view and is located at an angular separation of 5'.3 away from the target. Fortunately, the reference is of similar colour and brightness, which reduced the effect of differential colour extinction. For example, the magnitude differences (target minus reference) from the PPMXL[33] catalogue are $\Delta B = -0.46$, $\Delta R = -0.49$, $\Delta I = -0.5$. We observed both transits with the same slit mask and the red detector (MIT), which is a mosaic of two chips. We positioned the instrument field of view such that each detector imaged one source. The field of view was monitored without guiding interruptions during the full observing campaigns. To improve the duty cycle, we made use of the fastest available read-out mode (200 kHz, ~30s). During both nights, we ensured that the Longitudinal Atmospheric Dispersion Corrector (LADC) is in its neutral position, i.e. inactive.

During the first night, we used the dispersive element GRIS600B (hereafter blue and 600B), which covers the spectral range from 3600 to 6200 Å at a resolving power of R ~ 600. The field of view rose from an airmass 1.43 to 1.08 and set to an airmass of 1.16. The seeing oscillated around 0.5" during the first 3.5 hours and gradually increased to 1.2" at the end of the observation. We collected a total of 89 exposures for ~5h with an integration times adjusted between 120 and 230s.

During the second night, we exploited the dispersive element GRIS600RI (hereafter red and 600RI), which covers the range from 5400 to 8200 Å, in combination with the GG435 filter to isolate the first order. The field of view rose from an air mass of 1.23 to 1.08 and set at an airmass of 1.36. The seeing varied between 0.3" and 0.5" as measured from the cross-dispersion profiles of the spectra. We monitored WASP-96 and the reference star for ~5h20m and collected a total of 233 spectra with integration times between 30 and 80s.

## Calibrations and data reduction.

We performed data reduction and analysis using a customized IDL pipeline[22]. We started by subtracting a bias frame and by applying a flat field correction to the raw images. We computed a master bias and flat field by obtaining the median of 100 individual frames. Cosmic rays were identified and corrected following the routine detailed in ref. 34. We extracted 1D spectra using IRAF's APALL task. To trace the stars, we used a fit of a Chebyshev polynomial of two parameters. We performed background correction by subtracting the median background from the stellar spectrum for each wavelength, computed from a box located away from the spectral trace. We found that aperture radii of 14 and 12 pixels and sky regions 21: 72 and 23: 74 minimize the dispersion of the out-of-transit flux of the band-integrated white light curves for the blue and red observations, respectively.

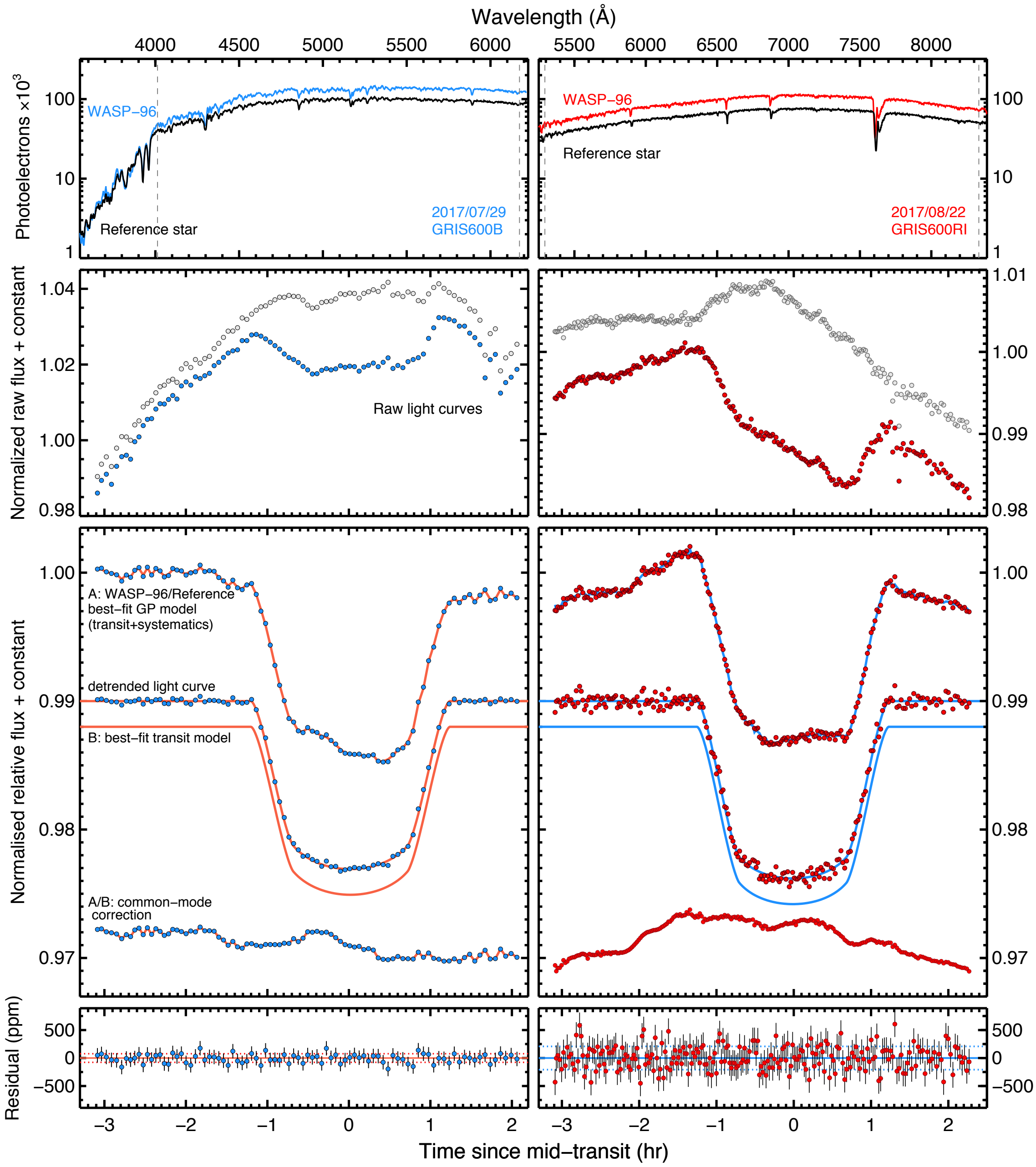

**Extended Data Figure 1 | VLT FORS2 stellar spectra and white-light curves.** Left and right column panels show the GRIS600B (blue) and GRIS600RI (red) data sets, respectively. **First row:** example stellar spectra used for relative spectrophotometric calibration. The dashed lines indicate the wavelength region used to produce the white-light curves. **Second row:** normalised raw light curves of both sources. **Third row:** Normalised relative target-to-reference raw flux along with the marginalized GP model (A), detrended transit light curve and model (B), and the common-mode correction (A/B), **Fourth row:** Best-fit light curve residuals and $1\sigma$ error bars, obtained by subtracting the marginalized transit and systematics models from the relative target-to-reference raw flux. The two light curve residuals show dispersion of 78 and 201 parts-per-million (ppm), respectively.

We performed a wavelength calibration of the extracted stellar spectra using spectra of an emission lamp, obtained after each of the two transit observations with a mask identical to the science mask, but with slit widths of 1". We established a wavelength solution for each of the two stars with a low-order Chebyshev polynomial fit to the centres of a dozen lines, which we identified by performing a Gaussian fit. To account for displacements during the course of each observation and relative to the reference star, we placed the extracted spectra on a common Doppler-corrected rest frame through cross-correlation. All spectra were found to drift in the dispersion direction to no more than 2.5 pixels, with instrument gravity flexure being the most likely reason.

Example spectra of WASP-96 and the reference star are shown in Extended Figure 1. We achieved typical signal-to-noise ratios (S/N) of 315 and 280 per pixel for the central wavelength of the blue grism and 313 and 257 for the red grism, respectively. We then used the extracted spectra to produce band-integrated white and spectroscopic light curves for each source and transit by summing up the flux along the dispersion axis in each bandpass.

**White light curve analysis.**

We produced white light curves from 4013 to 6173 and from 5293 to 8333 Å for the blue and red observations, respectively. We corrected the raw flux of the target by dividing by the raw flux of the reference star. This correction removes the contribution of Earth's atmospheric transparency variations, as demonstrated in Extended Data Figure 1. We modelled the white light transits and instrumental systematics simultaneously by treating the data as a Gaussian process (GP)[35-37]. We performed the GP analysis using the Python GP library *George*[38-41]. Under the GP assumption, the data likelihood is a multivariate normal distribution with a mean function $\mu$ describing the deterministic transit signal and a covariance matrix K that accounts for stochastic correlations (i.e. poorly-constrained systematics) in the data:

$$p(f\,|\theta,\gamma) = \mathcal{N}(\mu,\mathrm{K}),$$

where $p$ is the probability density function, $f$ is a vector containing the flux measurements, $\theta$ is a vector containing the mean function parameters, $\gamma$ is a function containing the covariance parameters, $\mathcal{N}$ is a multivariate normal distribution. We defined the mean function $\mu$ as follows:

$$\mu(t,\hat{t};c_0,c_1,\theta) = [c_0 + c_1\hat{t}]\mathrm{T}(t;\theta),$$

where $t$ is a vector of all central exposure time stamps in Julian Date (JD), $\hat{t}$ is a vector containing all standardized times, i.e. with subtracted mean exposure time and divided by the standard deviation, $c_0$ and $c_1$ describe a linear baseline trend, T(θ) is an analytical expression describing the transit and $\theta = (i, a/R_*, T_{mid}, R_p/R_*, u_1, u_2)$, where $i$ is the orbital inclination, $a/R_*$, is the normalized semi-major axis, $T_{mid}$ is the central transit time, $R_p/R_*$ is the planet-to-star radius ratio, $u_1$ and $u_2$ are the linear and quadratic limb darkening coefficients. To obtain an analytical transit model T, we used the formulae found in ref. 42. We fixed the orbital period to its value from ref. 13 and fitted for the remaining system parameters.

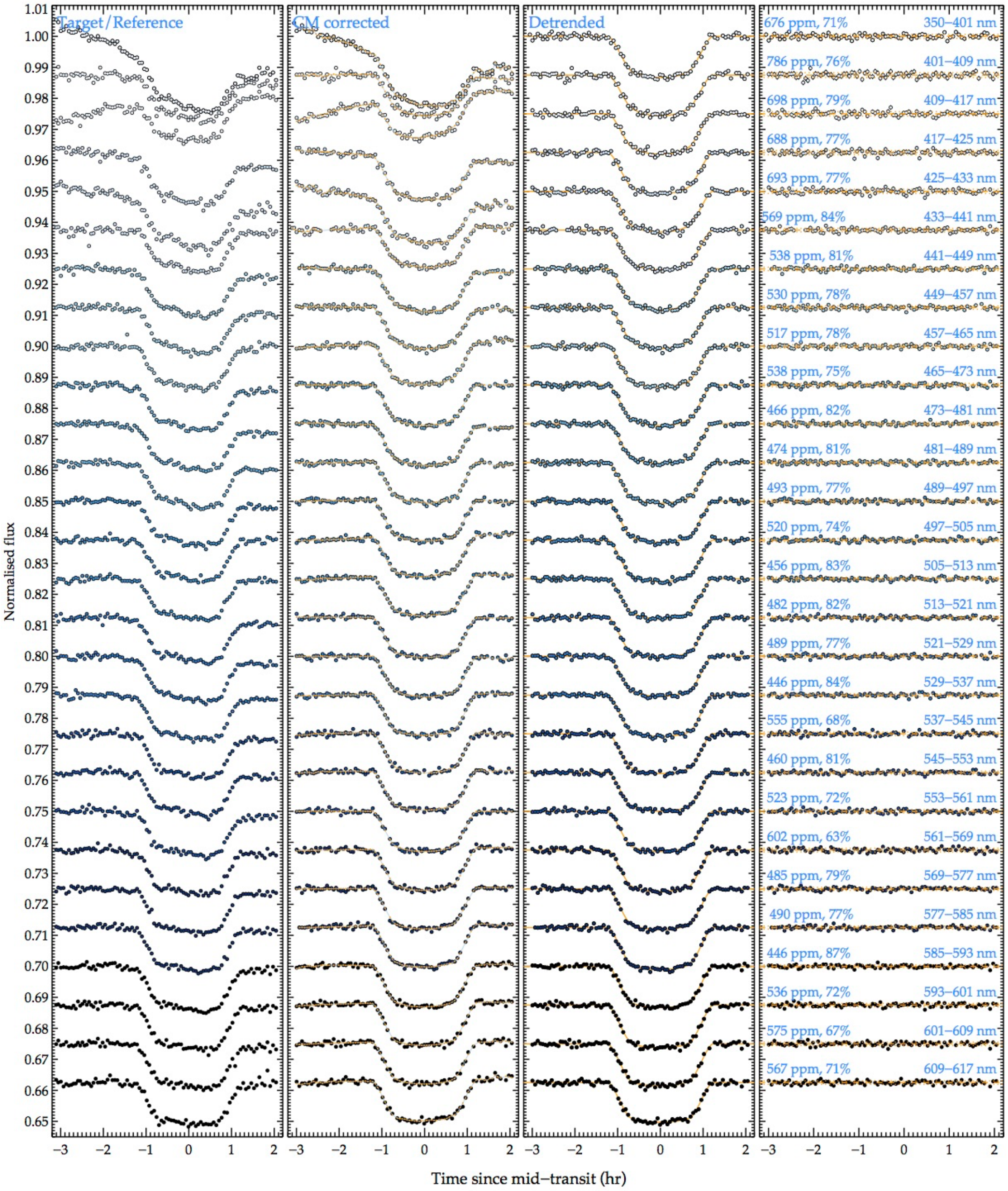

**Extended Data Figure 2 | Spectrophotometric light curves from grism 600B** offset by a constant amount for clarity**. First panel:** raw target-to-reference flux. **Second panel:** common-mode corrected light curves and the transit and systematics models, with the highest evidence. **Third panel:** detrended light curves and the transit model with the highest evidence. **Fourth panel:** residuals with the 1-σ error bars. The dashed lines indicate the median residual level, with dotted lines indicating the dispersion and the reached percentage of the theoretical photon noise limit (blue).

We accounted for the stellar limb-darkening by adopting the two-parameter $(u_1, u_2)$ quadratic law and computed the values of the coefficients using a three-dimensional stellar atmosphere model grid[43]. In these calculations, we adopted the closest match to the effective temperature, surface gravity and metallicity of the exoplanet host star found in ref. 13. The choice of a quadratic versus a more complex law, e.g. the four-parameter non-linear law of ref. 44 was motivated by the study of refs. 45 and 46, where the two-parameter law has been demonstrated to introduce negligible bias on the measured properties of transiting systems similar to WASP-96. In addition, the quadratic law requires much shorter computational time for the relevant transit light curve. We computed the theoretical limb-darkening by fitting the limb-darkened intensities of the 3D stellar atmosphere models, factored by the throughputs of the blue and red grisms.

The covariance matrix is defined as

$$\mathrm{K} = \sigma_i^2 \delta_{ij} + k_{ij},$$

where $\sigma_i$ are the photon noise uncertainties, $\delta_{ij}$ is the Kronecker delta function and $k_{ij}$ is a covariance function. We assumed the white noise term was the same for all data points and allowed it to vary as a free parameter, $\sigma_w$. For the covariance function, we choose to use the Matérn $\nu = 3/2$ kernel with the spectral dispersion and cross-dispersion drifts $x$ and $y$, respectively as input variables, FWHM, measured from the cross-dispersion profiles of the 2D spectra and the speed of the rotation angle $z$ (see Extended Data Figure 4). As with the linear time term, we also standardised the input parameters prior to the light curve fitting. We chose to use the dispersion and cross-dispersion drifts for both observations, and combined them with the FWHM for the blue data and the speed of the rotation angle for the red data, respectively. Our choice was justified based on the fact that those combinations of input parameters gave well-behaved residuals. The covariance function then was defined as:

$$k_{ij} = \mathrm{A}^2(1 + \sqrt{3}D_{ij}) \exp\left( - \sqrt{3}D_{ij}\right),$$

where A is the characteristic correlation amplitude and

$$D_{ij} = \sqrt{\frac{(\hat{x}_i - \hat{x}_j)^2}{\tau_x^2} + \frac{(\hat{y}_i - \hat{y}_j)^2}{\tau_y^2} + \frac{(\hat{z}_i - \hat{z}_j)^2}{\tau_z^2}}$$

where $\tau_x$, $\tau_y$ and $\tau_z$ are the correlation length scales and the hatted variables are standardised. We allowed parameters $X = (c_0, c_1, T_{mid}, i, a/R_*, R_p/R_*, u_1, u_2)$ and $Y = (\mathrm{A}, \tau_x, \tau_y, \tau_z)$ to vary and fixed the orbital period $P$ to its literature value[13]. We adopted uniform priors for $X$ and log-uniform priors for $Y$.

To marginalise the posterior distribution $p\ (\theta, \gamma |\ f) \propto p\ (f\ |\theta, \gamma)\ p\ (\theta, \gamma)$ we made use of the Markov-Chain Monte Carlo (MCMC) software package *emcee*[40]. We identified the maximum likelihood solution using the Levenberg–Marquardt least-squares algorithm[47] and initialised three groups of 150 walkers close to that maximum. We run groups one and two for 350 samples and the third group had 4500 samples. Before running for the second group we re-sampled the positions of the walkers in a narrow space around the

position of the best walker from the first run. This extra re-sampling step was useful because otherwise some of the walkers can start in a low likelihood area of parameter space and would require more computational time to converge. Transit models for each of the two observations computed using the marginalized posterior distributions are shown in Extended Data Figure 1 and the relevant parameter values are reported in Extended Data Table 1. We find residual dispersion of 79 and 203 parts-per-million for the blue and red light curves, respectively. Both values are found to be within 80% of the theoretical the photon noise limit.

We computed the weighted mean values of the orbital inclination and semi-major axis and repeated the fits. In the second fit we allowed only the planet-to-star radius ratio ($R_p/R_*$) and the two limb darkening coefficients ($u_1$ and $u_2$) to vary, while the orbital inclination and semi-major axis were held fixed to the weighted mean values and the central times were fixed to the values determined from the first fit.

**Spectroscopic light curve analysis.**
We produced spectroscopic light curves by summing the flux of the target and reference star in bands with a width of 160Å. The sodium D lines at 5890 and 5896Å fall inside the spectral range of each grism, within their overlapping region from 5300 to 6200Å. We centred the set of bins for each night on the sodium line, which gave the advantage of obtaining two radius measurements identical in wavelength coverage within that overlapping region. We merged two pairs of bins, covering the $O_2$ A and B bands from 7594 to 7621 and from 6867 to 6884 Å, respectively to increase the signal to noise ratio of the corresponding light curves. The very first band in the blue grism was also enlarged with the same motivation. With these customizations at hand we produced a total of 63 light curves.

**Common mode factors**
The FORS2 spectroscopic light curves are known to exhibit wavelength independent (common mode) systematic, as demonstrated from our Comparative Transmission Spectroscopy studies[22,32] and other FORS2 results[48-50]. This makes the instrument outstanding for transmission spectroscopy, with enormous potential to explore the diversity of exoplanet atmospheres. We established the wavelength independent systematic using the band-integrated white light curves for each of the two nights. We simply divided the white-light transit light curves by a transit model. We computed the transit model using the weighted mean values of the orbital inclination and normalized semi-major axis from both nights and assumed the central times found from the white light analysis. The values for the relative radius and the limb darkening coefficients were identified by repeating the GP fit where the transit central time, orbital inclination and semi-major axis were held fixed to the weighted-mean values. The fitted relative radii and limb-darkening coefficients are reported at the end of Extended Data Table 1. The common mode factors for each night are shown in Extended Data Figure 1 along with a schematic explanation of the full white-light curve analysis.

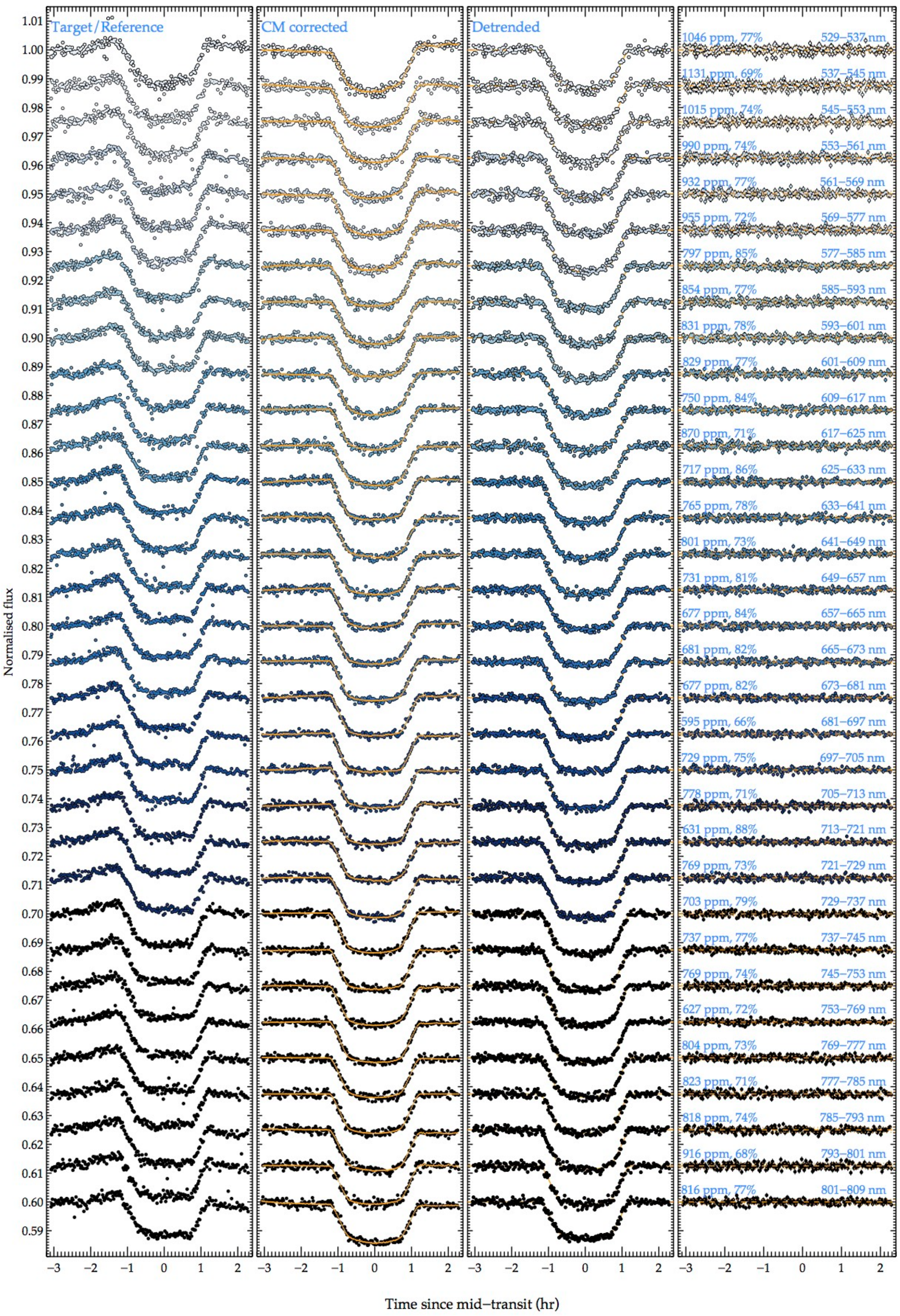

**Extended Data Figure 3 | Same as Extended Data Figure 2 but for grism 600RI**.

**Spectroscopic light curve fits**

We modelled the spectroscopic light curves using a two-component function that takes into account the systematics and transit simultaneously. The transit model was computed using the analytical formulae of ref. 42, as for the white light curve, but we allowed in the fits only the relative planet radius $(R_p/R_*)$ and the linear limb darkening coefficient $(u_1)$ to vary. Fitting only for the linear limb-darkening coefficient is a standard practice for ground-based observations that has been proven to generally perform well[51-56]. Similar to our WASP-39b study with FORS2[22], we also fit for both limb-darkening coefficients and found that the uncertainty of $u_2$ is large and consistent with the theoretical prediction. We interpret this as an indication for insufficient constraining power of the data for the non-linear coefficient. However, as the transmission spectra did not significantly change we chose to fix $u_2$ and to fit only for $u_1$. Prior to fitting the spectroscopic light curves, we removed the common mode factors from each night by dividing each of the spectroscopic light curves to the corresponding common-mode light curve of the same night. We accounted for the systematics using a low-order polynomial (up to a second degree with no cross terms) of dispersion and cross-dispersion drift, air mass, FWHM variations and the rate of change of the rotator angle (Extended Data Figure 4). We produced all possible combinations of detrending variables and performed separate fits with each combination with the systematics function included in the two-component model. For each attempted function, we computed the Akaike Information Criterion (AIC)[57] to estimate statistical weight of the model depending on the number of degrees of freedom. We marginalized the resulting relative radii and linear limb-darkening coefficient following ref. 58. We chose to rely on the AIC instead of other information criteria, e.g. the Bayesian Information Criterion (BIC)[59], because the AIC selects more complex models, resulting in more conservative error estimates. We note that a marginalization over multiple systematics functions relies on the assumption of equal prior weights for each tested model. This assumption is valid for simple polynomial expansions of basis parameters, as the ones in our study. We found systematics models, parameterized with a linear air mass, dispersion drift and FWHM terms to result in the highest evidence.

Prior to each fit, we set the uncertainties of each spectrophotometric channel to values that are based on the expected photon noise with additional component from readout noise. We determined best-fit models using a Levenberg-Marquardt least-squares algorithm and rescaled the uncertainties of the fitted parameters with the dispersion of the residuals. All residual outliers larger than $3\sigma$ were excluded from the analysis. We found that for each spectroscopic light curve $\leq 3$ data points were removed. We also assessed the levels of correlated residual red noise, following the methodology of ref. 60 by modelling the binned variance with $\sigma^2 = (\sigma_w)^2/N + (\sigma_r)^2$ relation, where $\sigma_w$ is the uncorrelated white noise component, N is the number of measurements in the bin and $\sigma_r$ is the red noise component. We find white and red noise dispersion in the range from 400 to 1000 and from 20 to 80 ppm, respectively.

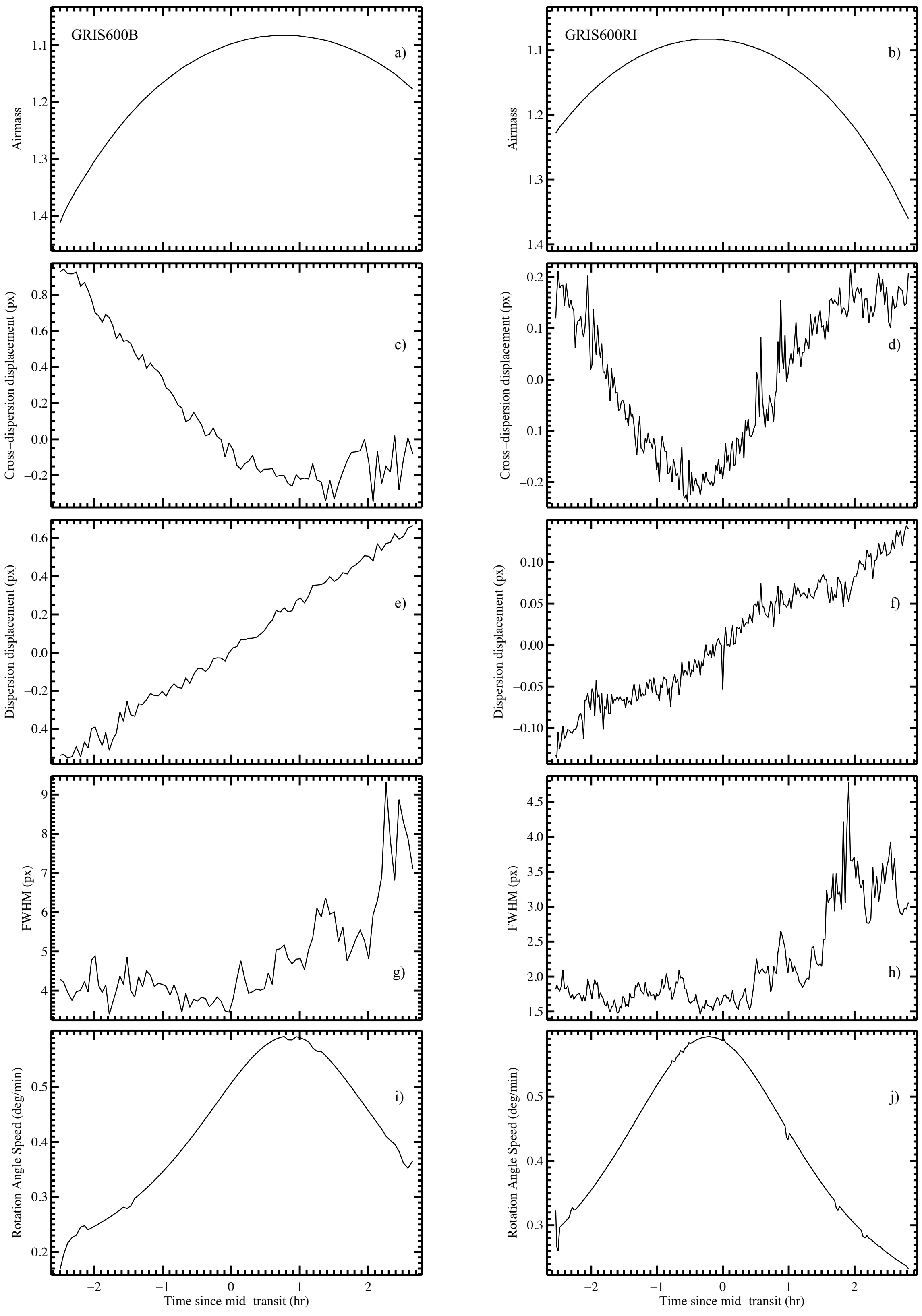

**Extended Data Figure 4 | Light curve auxiliary variables.** Shown are air mass (a, b), drifts along the cross-dispersion (c, d) and dispersion axes (e, f), full width at half maximum (g, h) and the rate of change of the rotation angle (i, j) of the VLT FORS2 observations. Left and right columns refer to the GRIS600B and GRIS600RI observations, respectively.

The measured wavelength dependent relative radii and corresponding light curves are plotted in Figure 1 and Extended Data Figures 2, 3, 4, 5, Table 2 and comprise the transmission spectrum of WASP-96b. We used the overlapping wavelength region to combine the two observations together and computed the weighted mean of both datasets. We detect a marginally significant 1.4$\sigma$ difference in the transit depths of the light curves from the two observations of $(7.2 \pm 5.2) \times 10^{-4}$. This level of variation is consistent with the photometric variability, associated with the active regions on the star surface with variability of $\sigma < 9.2 \times 10^{-4}$ (ref. 13).

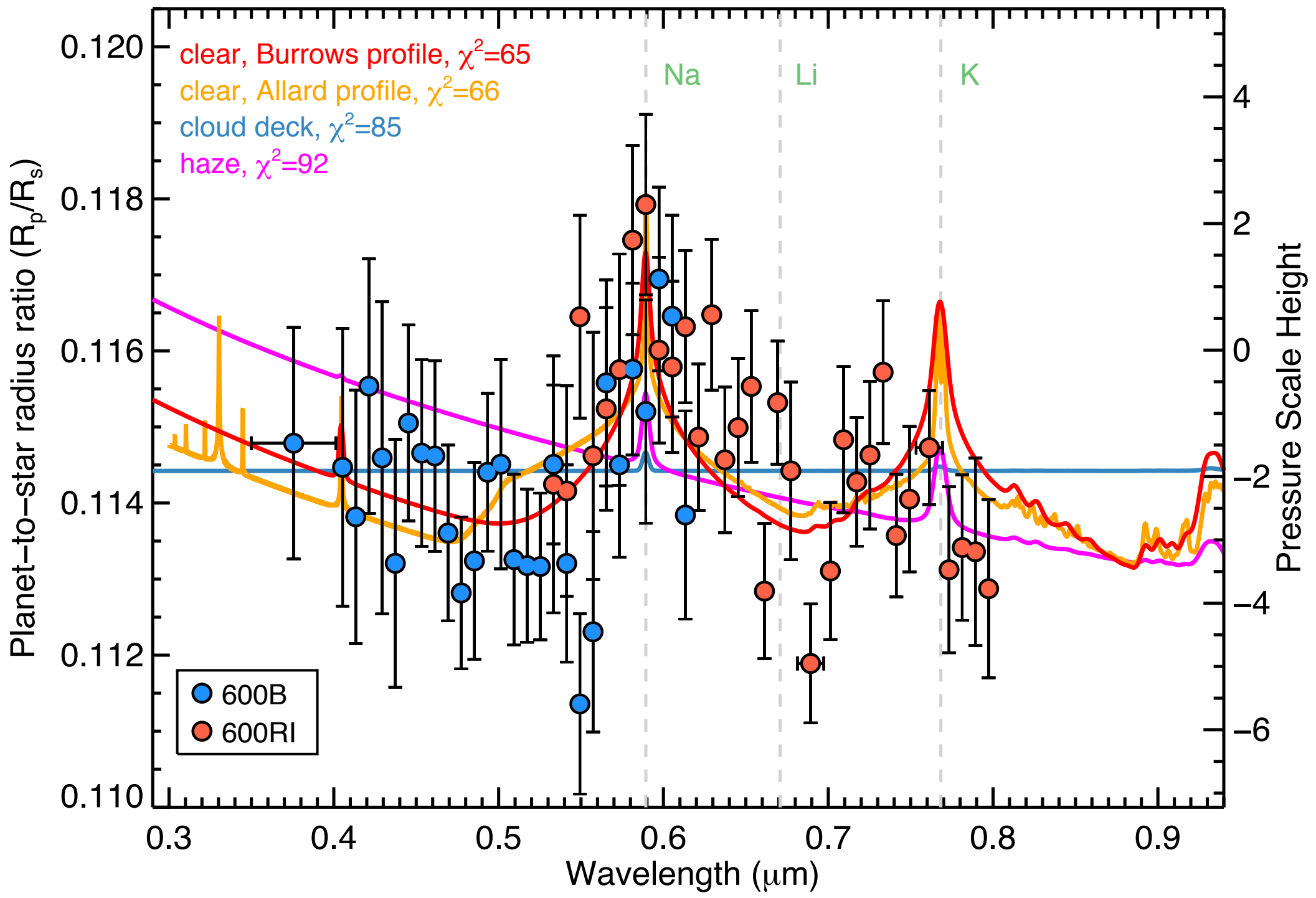

**Extended Data Figure 5 | Transmission spectrum of WASP-96b.** Indicated are the relative radius measurements from grism 600B (blue dots) and 600RI (red dots) along with the 1$\sigma$ uncertainties, compared to the same set of models as in Figure 1.

**Modelling of the atmosphere of WASP-96b.**

First we compare the observed transmission spectrum to the models based on ref. 16, which included a self-consistent treatment of radiative transfer and chemical equilibrium of neutral and ionic species. Chemical mixing ratios and opacities were computed assuming solar metallicity and local chemical equilibrium, accounting for condensation and thermal ionization but not photoionization[61-63]. Transmission spectra were also calculated using 1D T-P profiles for the dayside, as well as an overall cooler planetary-averaged profile.

In addition to the atmospheric models of a clear atmosphere, a simplified treatment of the scattering and absorption have been incorporated in those models to simulate the effect of small particle haze aerosols and large particle cloud condensates at optical and near-infrared wavelengths. In the case of haze,

Rayleigh scattering opacity ($\sigma = \sigma_0(\lambda/\lambda_0)^{-4}$) has been assumed with a cross-section which was 1,000× the cross-section of molecular hydrogen gas ($\sigma_0 = 5.31 \times 10^{-27}\text{cm}^2$ at $\lambda_0 = 3500\text{Å}$; ref. 61). To include the effects of a flat cloud deck we included a wavelength-independent cross-section, which was a factor of 100× the cross-section of molecular hydrogen gas at $\lambda_0 = 3500\text{Å}$; (see Fig. 1).

Similar to our previous studies, we obtained the average values of the models within the wavelength bins and fitted these theoretical values to the data with a single parameter responsible for their vertical offset[5,22,34]. The $\chi^2$and BIC statistic quantities were computed for each model with the number of degrees of freedom for each model determined by $\nu = N - m$, where $N$ is the number of measurements and $m$ the number of free parameters in the fit.

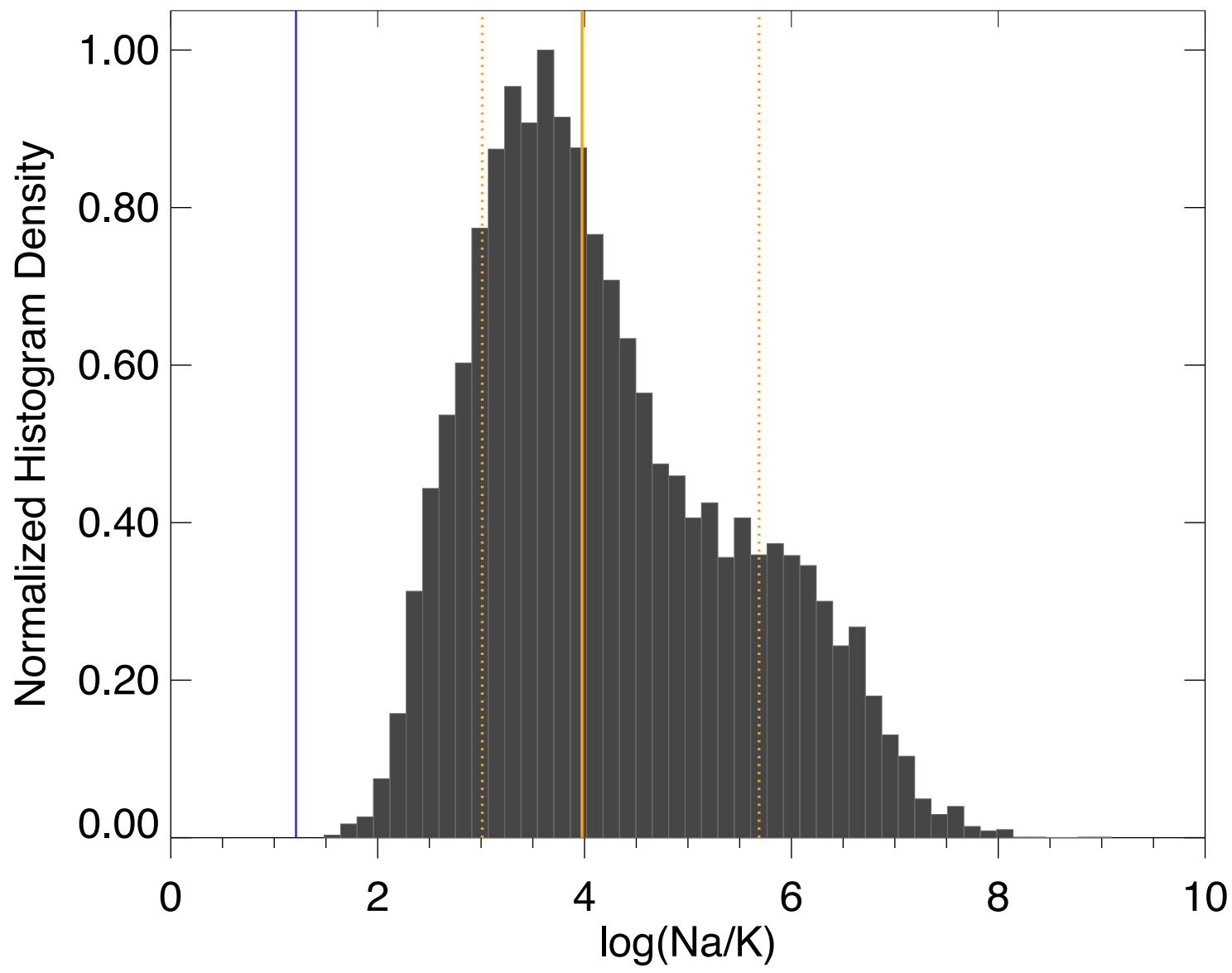

**Extended Data Figure 6 | Na/K ratio** Histogram of the marginalized posterior distribution of the Na to K ratio for WASP-96b. Shown are the median and 1$\sigma$ levels with orange continuous and dotted lines, respectively. The solar value is indicated with the blue continuous line.

We also compared the observed transmission spectrum around the Na D line from 4500 to 6800 Å with cloud free atmosphere models, assuming two individual shapes of the pressure broadened profile. The first profile had the shape detailed in ref. 3 and the second line shape from the formalism of ref. 18, 26. We find no statistically significant difference between the two with $\chi^2 = 22$ and $\chi^2 = 30$ for 35 d.o.f. Further observations at higher signal to noise and resolution will be necessary to distinguish between the two line-wing shapes.

**Retrieval analysis**

We performed a retrieval[8,9,64] analysis using the 1D radiative-convective equilibrium ATMO model[24,65-68]. The code includes isotropic multi-gas Rayleigh scattering, $H_2$-$H_2$ and $H_2$-He collision-induced absorption, as well as opacities for all major chemical species taken from the most up-to-date high-temperature sources, including: $H_2$, He, $H_2O$, $CO_2$, CO, $CH_4$, Na, K, Li, Rb and Cs [66,69]. Hazes and clouds are respectively treated as parameterized enhanced Rayleigh scattering and grey opacity[70], which is similar to the approach of ref. 16 and is detailed in ref. 69. ATMO uses the correlated-k approximation with the random overlap method to compute the total gaseous mixture opacity, which has been shown to agree well with a full line-by-line treatment[66].

We first performed a free retrieval analysis, allowing a fit to the abundances of Na, K and Li, cloud and haze opacity, the planet's radius and atmospheric temperature. Na, K and Li were included as these elements are expected to add significant opacity in the wavelength region of our observations. We assumed that these gasses are well-mixed vertically in the atmosphere.

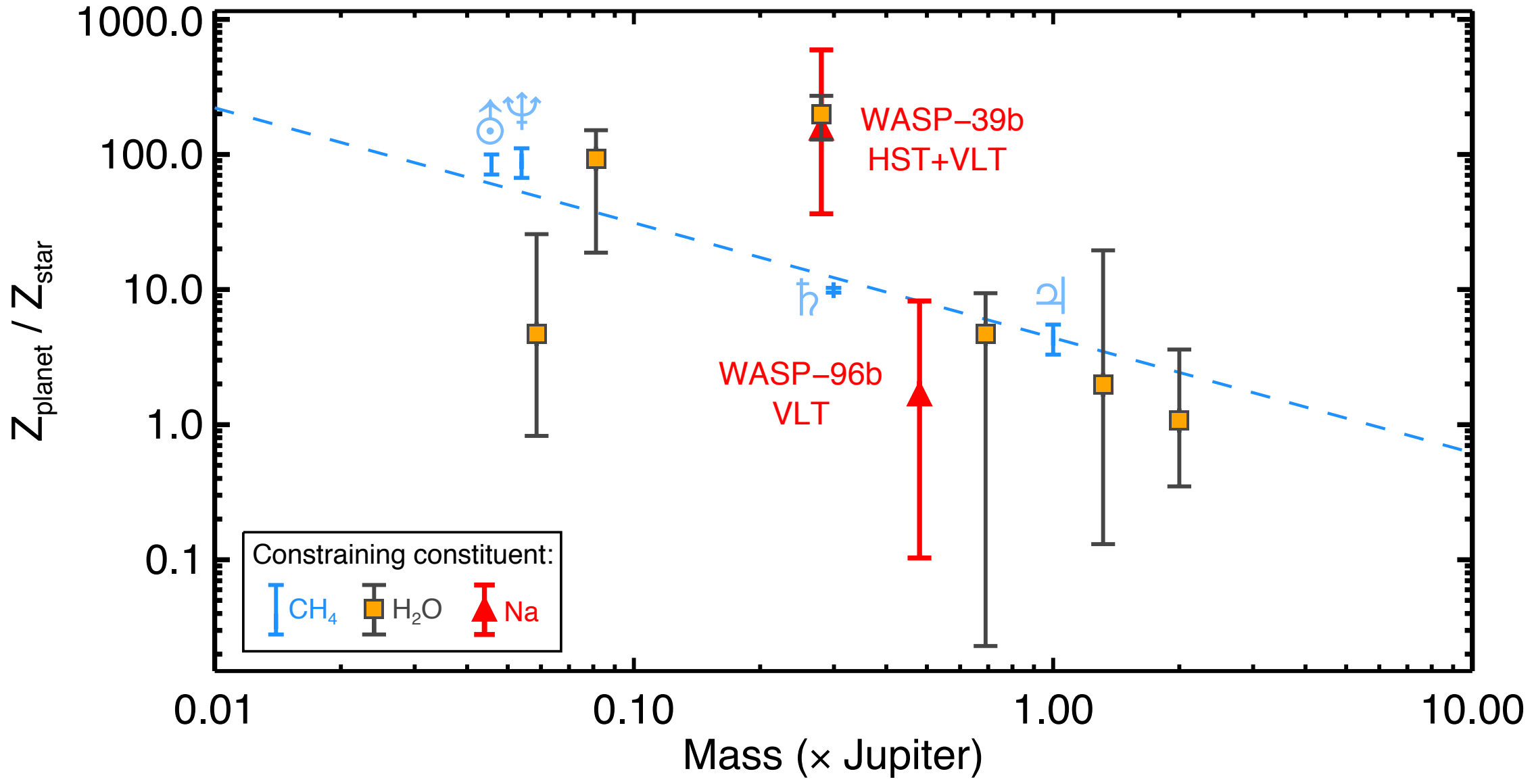

**Extended Data Figure 7 | Heavy element enrichment of exoplanets relative to their stars as a function of mass.** Plotted are the solar system planets (blue bars) and gas-giant exoplanets (grey and orange symbols). Each error bar represents the $1\sigma$ uncertainty. The blue line indicates a fit to the Solar System gas giants (pale blue symbols indicate Solar System planets).

For all free parameters in our model we adopted uniform priors with the following ranges: $250 - 3000$K for the temperature, $1 - 1.5R_J$ for the radius, $10^{-10}$ to $10^3$ for the cloud and $10^{-10}$ to $10^3$ for the haze opacity. For the mixing ratios of chemical species other than H and He we adopted uniform priors between $-2$ and 12. Our retrieval analysis proceeded by first identifying the minimum $\chi^2$ solution using nonlinear least-squares optimization and then marginalizing over the posterior distribution using differential-evolution Markov chain Monte Carlo[71]. A total of 12 chains for 30,000 steps each, were ran until the Gelman-Rubin statistic for each free parameter was within 1% of unity, showing that the chains were well-mixed and had reached a steady state. We discarded a burn-in phase from all chains corresponding to the step at which all chains had found a $\chi^2$ below the median $\chi^2$ value of the chain. Finally, we combined the remaining samples into a single chain, forming our posterior distributions. We summarize the results from the free retrieval in Figure 2.

Elemental abundances of Na and Li were obtained for the host star from high-resolution optical spectroscopy with typical S/N of ~100:1, as detailed in ref. 13. Because the potassium resonance doublet is located outside of the red limit of the high-resolution spectra, it was not possible to obtain abundance measurement.

The best-fitting retrieved spectrum suggests a cloud-free atmosphere at the limb with precise sodium abundance consistent with the solar value at $\sim 1\sigma$. The spectrum shows no definitive evidence of the potassium feature. A measurement at ~0.73$\mu$m shows a slightly larger radius, though it is only $\sim 1.8\sigma$ above our best-fit model. Our retrieval analysis shows a marginalized posterior distribution bounded only on the upper bound with median value consistent with the solar value at the $\sim 2\sigma$ confidence (Figure 1 and 2). This most likely is a consequence of the limited wavelength range, covering only ~2/3 of the potassium feature, as well as the diluting effect of the $O_2$-A band, which partially covers this wavelength regime. The limited constraining power of the K feature allows for a three-case scenario for the abundance of K in the atmosphere of WASP-96b. In both cases the atmosphere of the planet is clear at the limb with sodium at solar abundance. In the first scenario, the abundance of potassium is sub-solar, leading to a missing potassium feature in the transmission spectrum. In chemical equilibrium, Na and K can form condensates (e.g $Na_2S$ and KCl) at temperatures lower that $\sim 1300$ and $1000$ K, respectively. However, given that the retrieved temperature in the region probed by our observation lies between $\sim 1300$ to $1700$ K this is unlikely.

In the second scenario, the potassium can be ionized by the UV radiation of the host star, which would also lead to missing potassium line cores. The pressure-broadened line wings would still be present, because they would originate from deeper layers, where the UV radiation would be able to penetrate. We estimate the relative sodium-to-potassium ratio from our free retrieval analysis, finding highly uncertain value, which prevents definitive conclusion (Extended Data Figure 6). Future observations of multiple transits could help resolve the two-case scenario.

A third scenario would require stratified atmosphere consisting of a low-altitude potassium layer followed by a cloud cover on top, above which is located a clear atmosphere consisting sodium at solar abundance. The sodium layer would need to be deep enough to allow for pressure-broadened line wings in the transmission spectrum.

We note a large degeneracy between the aerosol clouds/hazes and the Na abundance, as higher cloud opacity levels can be fit with increased Na abundances. This degeneracy can be seen in the posterior distribution of the retrieval (see Extended Data Figure 8), and the marginalized posterior distributions are shown in Figure 2 (i.e. the distribution shown in Extended Data Figure 8 integrated along each axis). The degeneracy is accounted for in our retrieval modelling by marginalizing the Na abundance over the other fit model parameters, which includes the effects of clouds and hazes. An aerosol-free model where the near-UV transmission spectra is dominated by $H_2$ Rayleigh scattering helps determine the lower limit to the Na abundance. The upper limit for the Na abundance is sensitive to the line profile shape, as very high aerosol opacity levels probe significantly lower pressure levels, which affects the line profile. For example, for a clear atmosphere the transmission spectra at 4000 Å probes pressures of 20 mbar and is dominated by molecular hydrogen at that wavelength. In a model where the haze opacity is 1000x stronger than molecular hydrogen (the pink haze model shown in Figure 1), the pressure probed is ~1000x lower or ~0.02 mbar. These lower pressures affect the calculation of the pressure-broadened sodium line profile. As described in ref. 3, using semi-classical impact theory the half-width of the Lorentzian sodium line core is determined by the effective collision frequency, which itself is a product of the $H_2$- perturber density (among other factors). We find that in the described scenario with very high sodium abundances and high cloud/haze opacities, very low pressures (such as 0.02 mbar) are probed and the sodium line profile becomes too narrow to be able to fit the wide profile as seen in the data, even at arbitrarily high abundances.

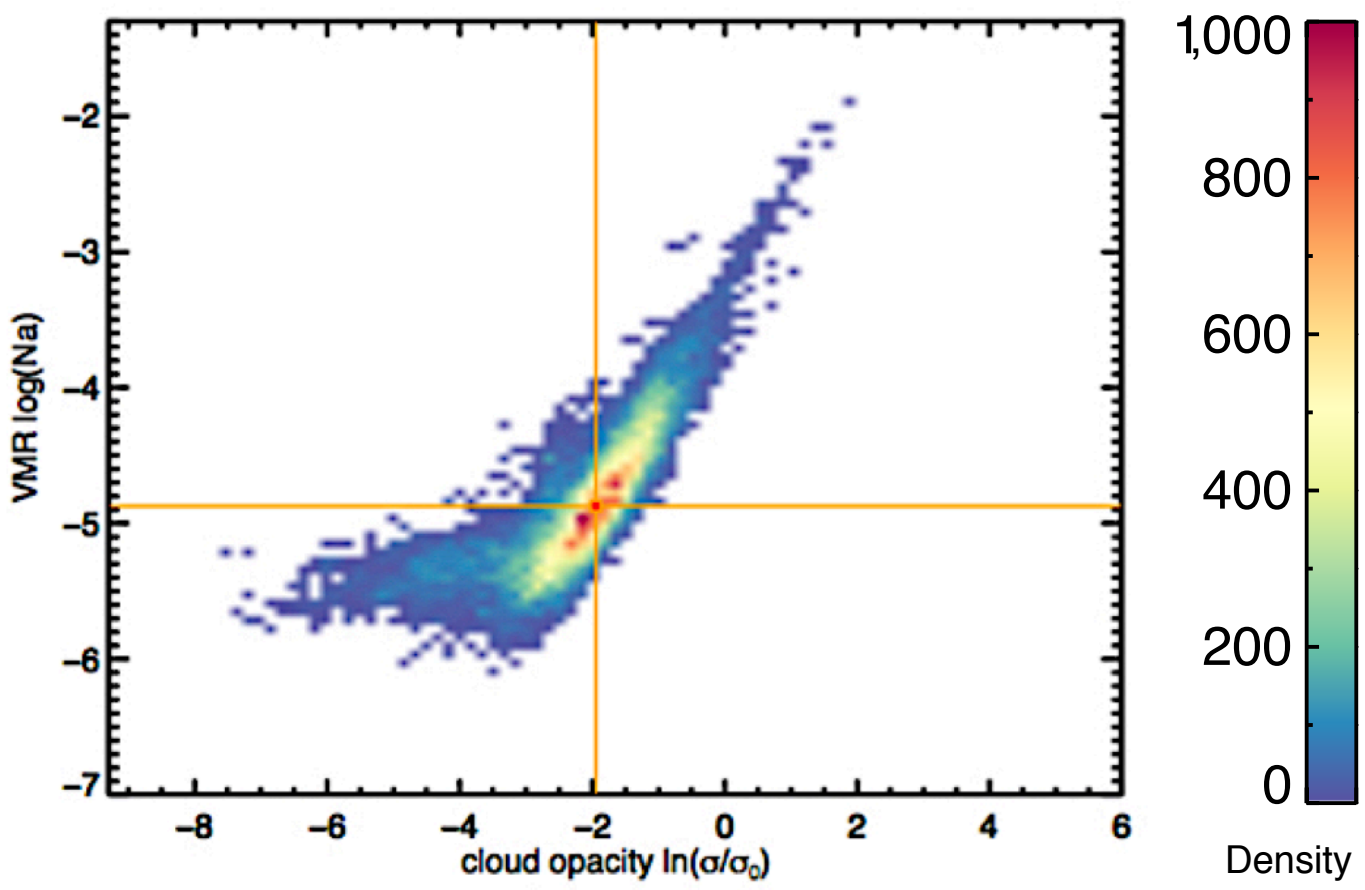

**Extended Data Figure 8 | Posterior distribution of the retrieved cloud opacity versus the sodium abundance.** VMR, vertical mixing ratio. The median and 1$\sigma$ measured parameters are indicated with continuous lines and the red dot marks the intercept for clarity. The colour scale shows the normalized density of the samples in the MCMS run.

We acknowledge that the 1D model does not account for horizontal advection which may affect the composition. This can break the chemical equilibrium conditions, leading to horizontally-constant non-equilibrium abundances of various species e.g., $CH_4$ (ref. 72), which could potentially affect the observed transmission spectrum. However, an analysis of such effects is beyond the scope of the present letter.

The presence of sodium and potassium in the atmospheric spectra of irradiated gas giant exoplanets has proved puzzling, with some of the spectra showing both or either of the two features without a clear trend with the planetary properties[5]. Primordial abundance variation along with atmospheric processes, such as condensation, photochemistry and photoionisation have been hypothesized to be responsible for the observed alkali variation. Elemental abundance constraints with ground-based instruments such as FORS2 could help identify the role of some of those processes from statistically large samples of exoplanet spectra.

Exoplanet atmospheric metallicities have been estimated using scaling relation of atmospheric metallicity with abundance of detected absorption features. Currently, such estimations have been obtained using the 1.4 µm water absorption feature. Recently, ref. 73 have obtained such measurement from a combined HST + VLT spectroscopy. We performed a retrieval analysis assuming chemical equilibrium in the atmosphere and determine the metallicity of WASP-96b's atmosphere, finding consistency with the metallicity of the host star and the mass-metallicity trend established for solar system giant planets (Figure 3). We acknowledge that such estimations can largely be inaccurate due to the scaling metallicity assumption with elemental abundance and round-off the precision of the WASP-96b's metallicity to one significant figure. The current census of exoplanet metallicity measurements exhibits a large scatter across the full mass-metallicity diagram, which hampers definitive conclusion regarding a trend. Similar to ref. 74 we plot the relative planet to star heavy element enrichment as a function of the planet mass instead of the planet metallicity alone (Extended Data Figure 7). We convert $Z_{\mathrm{p}}/Z_{\odot}$ to $Z_{\mathrm{p}}/Z_{*}$ assuming a scaling approximation with the parent star iron abundance of the form $Z_{\mathrm{p}}/Z_{*} = 10^{[\mathrm{Fe/H}]}$, where $Z_{\odot} = 0.014$ and propagate the uncertainty of [Fe/H]. It remains a future work to improve the statistics on the mass versus metallicity diagram with additional precise measurements. We therefore, consider it pre-mature to claim hypothetical trends or emerging patterns and relevant interpretation.

**Code availability.** Publicly available custom codes were used for the Gaussian process modelling george (http://dfm.io/george/current/user/gp/) and emcee code (http://github.com/ dfm/emcee). The MCMC retrieval analyses were performed using the publicly available package exofast (http://astroutils.astronomy.ohio-state.edu/exofast). The ATMO code used to compute the atmosphere models is currently proprietary. We have opted not to make the customized IDL codes used to produce the spectra publicly available owing to their undocumented intricacies.

**Data availability.** The data is within standard proprietary period of one year and will become publicly available on ESO archive (July and August) in 2018. Reduced data products and models used in this study are available Supplementary Information.